\documentstyle[prl,aps,epsf]{revtex}
\begin{document}

\title{ Persistence of Randomly Coupled Fluctuating Interfaces}
\author{Satya N. Majumdar $^{1,2}$ and Dibyendu Das $^{3}$}
\address{
{\small $^1$Laboratoire de Physique Th\'eorique (UMR C5152 du CNRS), Universit\'e Paul
        Sabatier, 31062 Toulouse Cedex. France}\\
{\small $^2$Laboratoire de Physique Th\'eorique et Mod\`eles Statistiques,
        Universit\'e Paris-Sud. B\^at. 100. 91405 Orsay Cedex. France}\\
{\small $^3$ Department of Physics, Indian Institute of Technology Bombay, Powai, Mumbai 400076, 
India}\\}

\maketitle

\date{\today}

\begin{abstract}
We study the persistence properties in a simple model of two coupled interfaces
characterized by heights $h_1$ and $h_2$ respectively, each growing over a $d$-dimensional
substrate. The first interface evolves independently of the second 
and can correspond to any generic growing interface, e.g., of the Edwards-Wilkinson
or of the Kardar-Parisi-Zhang variety. The evolution of $h_2$, however, is
coupled to $h_1$ via a quenched random velocity field. In the limit
$d\to 0$, our model reduces to the Matheron-de Marsily model in two dimensions. For
$d=1$, our model describes a Rouse polymer chain in two dimensions 
advected by a transverse velocity field. We show analytically that
after a long waiting time $t_0\to \infty$, the stochastic process
$h_2$, at a fixed point in space but as a function of time, becomes
a fractional Brownian motion with a Hurst exponent, $H_2=1-\beta_1/2$,
where $\beta_1$ is the growth exponent characterizing the first  
interface. The associated persistence exponent is shown to be
$\theta_s^2=1-H_2=\beta_1/2$. These analytical results
are verified by numerical simulations.
\end{abstract}

\section{Introduction}
The survival probability $P(t)$ that a stochastic process
$X(t)$ does not cross zero upto time $t$ is a quantity of long standing interest in
probability theory and with many practical applications\cite{BL}. The derivative
$F(t) = -dP/dt$ is called the first-passage probability\cite{Redner}. This subject
has seen a resurgent interest over the last decade in the context of many body 
nonequilibrium systems where the stochastic process $X(t)$ denotes a local
time dependent field in a spatially extended evolving system. For example, 
in the case of the Ising model evolving under the Glauber dynamics, the relevant stochastic 
process $X(t)$ is the spin $s_i(t)$ at a fixed site $i$ as a function of time $t$ and
$P(t)$ then denotes the probability that the spin $s_i$ does not 
flip upto time $t$\cite{BDG}. In this context, the survival probability $P(t)$ thus measures the
`persistence' of a local field to remain in its initial state. In many of these
nonequilibrium systems, the persistence has been found to decay as a power law
at late times, $P(t)\sim t^{-\theta}$. The exponent $\theta$ is called the 
persistence exponent and has been a subject of much theoretical, numerical
and experimental studies in recent times\cite{Review}. The exponent $\theta$  
is very hard to calculate analytically even in simple systems such as the linear diffusion 
equation starting from random initial conditions\cite{diffusion}. The reason
for this difficulty can be traced back to the fact that the spatial
interactions in these extended systems makes the local stochastic
field $X(t)$ a `non-Markovian' process in time\cite{Review}.

Apart from these pure systems, persistence has also been studied in systems
with quenched disorder\cite{NS,Jain}. In presence of disorder, the exact
calculation of persistence $P(t)$ is nontrivial even for a single particle 
without any spatial interaction, though some analytical results have been obtained 
recently. For example, the asymptotic results for the persistence of a particle
moving in a random Sinai potential in one dimension have been obtained both in 
the case of a vanishing external field\cite{CD,IR,FDM} and also for nonzero external
field\cite{MC}. Another solvable example is the persistence of a single particle
advected by a layered random velocity field\cite{Redner1,SM,BG}.    

The purpose of this paper is to present analytical results for the persistence
$P(t)$ in a system with both spatial interaction and disorder. Our system 
consists of two growing $(d+1)$-dimensional interfaces where $d$ refers
to the dimension of the substrate on which the surfaces grow and $1$ refers
to the time $t$. The two interfaces are characterized by their heights
$h_1({\bf r}, t)$ and $h_2({\bf r}, t)$ respectively. In our model, while the height
$h_1$ of the first interface evolves independently of $h_2$, the evolution
of $h_2$ is driven by a velocity that is a random quenched function of
$h_1$. This random velocity represents the disorder in the system.
The model is detailed in Section II.  

There are two motivations for studying this model: 

\vspace{0.2cm}

(i) We will show later in Section III that our model corresponds to different well
known physical systems for different values of the spatial dimension $d$.
For example, in the limit $d\to 0$, our model reduces to the 
Matheron-de Marsily model where one studies the motion of a single Brownian particle 
in a $2$-dimensional plane in presence of a transverse random
velocity field\cite{MdM}. On the other hand, for $d=1$, we will show
that our model describes the evolution of a Rouse polymer chain\cite{Rouse}
(a Rouse chain consists of a set of beads or monomers connected by harmonic springs) 
moving in a $2$-dimensional plane and advected by a transverse random velocity field.
While the transport properties in this latter model are well understood\cite{OB,WL,JOB},
the persistence properties have not been studied before. The analytical results
obtained in this paper for the persistence in our model for general $d$ will apply to these 
models in the limiting cases $d=0$ and $d=1$.

\vspace{0.2cm}

(ii) The persistence properties of a single interface have been studied 
extensively both theoretically\cite{Krug,Krug1,MB1,M0,M1,M2,M3,Krug2} and more
recently they have been measured experimentally in a system of fluctuating
steps on crystal surfaces\cite{PE}. In the experimental system, there are many
step edges, each corresponding to a single interface. If the step edges
are sufficiently separated from each other, they can be considered as independently
growing interfaces. However, in general, there will always be an interaction, albeit
weak, between two step edges and their motions will be coupled. Hence it
is important to study persistence in models where the interface heights are coupled.
Motivated by this observation, we study here the persistence in a simple model of 
coupled interfaces. While this model, by no means, is a true representative of the 
actual experimental situation, the advantage is that it represents a minimal model 
with coupling for which one can calculate the persistence properties analytically.

The paper is organized as follows. In Section II, we define the model precisely
and state our main results. In Section III we consider the two limiting cases
$d\to 0$ and $d=1$ where our model reduces to two well known models. 
In Section IV we point out our strategy to compute
the persistence exponents analytically. In Section V, we map the relevant 
stochastic process into a fractional Brownian motion (fBm) and then using a well known
first-passage property of fBm, we calculate the persistence exponents analytically. 
Section VI considers
a generalization to other types of interfaces. Section VII contains details
of numerical results and finally we conclude with a summary and outlook in Section VIII.  

\section{The Model and the Main Results}

We consider two interfaces characterized by their heights $h_1({\bf r}, t)$ and $h_2({\bf r},t)$
growing on a $d$-dimensional substrate according to the following evolution equations
\begin{eqnarray}
\frac{\partial h_1}{\partial t} &=& \nabla^2 h_1 + \eta_1({\bf r}, t) \label{h1evol1} \\
\frac{\partial h_2}{\partial t} &=& \nabla^2 h_2 + v\left(h_1\left({\bf r},t\right)\right) 
+\eta_2({\bf r}, t),
\label{h2evol1}
\end{eqnarray}
where $\eta_1$ and $\eta_2$ represent the thermal Gaussian noises that are uncorrelated with each other, 
each has zero mean and their correlators are given by
\begin{eqnarray}
\langle \eta_1({\bf r_1}, t_1)\eta_1({\bf r_2}, t_2)\rangle &=& [4\pi a^2]^{-d/2}\,
e^{-|{\bf r_1}-{\bf r_2}|^2/{4a^2}}\, \delta(t_1-t_2) \label{eta1} \\
\langle \eta_2({\bf r_1}, t_1)\eta_2({\bf r_2}, t_2)\rangle &=& [4\pi a^2]^{-d/2}\,
e^{-|{\bf r_1}-{\bf r_2}|^2/{4a^2}}\, \delta(t_1-t_2) \label{eta2}
\end{eqnarray}
where $a$ denotes the range of the correlator and serves as a short distance cut-off. For $d<2$ 
where the interface roughens with time, the short distance cut-off plays
no important role and one can safely take the limit $a\to 0$ and replace the correlators
in Eqs. (\ref{eta1}) and (\ref{eta2}) by delta functions, i.e., 
$[4\pi a^2]^{-d/2}\,e^{-|{\bf r_1}-{\bf r_2}|^2/{4a^2}}\to \delta({\bf r_1} -{\bf r_2})$. However, for $d>2$, 
the Eqs. (\ref{h1evol1})
and (\ref{h2evol1}) exhibit ultra-violet divergences and one needs to keep a finite $a$
in order that $\langle h_1^2 \rangle$ and $\langle h_2^2\rangle$ remain finite in the
$t\to \infty$ limit.

The first interface $h_1$ evolves freely according to
Eq. (\ref{h1evol1}) which is precisely the celebrated Edwards-Wilkinson equation\cite{EW}
in $(d+1)$ dimensions. On the other hand, the evolution of the second interface $h_2$ in Eq. (\ref{h2evol1}),
in addition to having the Laplacian and the noise term, is
coupled to the height $h_1$ of the first interface via the quenched random velocity
$v\left(h_1({\bf r},t)\right)$ which is also considered to be a Gaussian with the following
moments
\begin{eqnarray}
{\overline {v(h_1)}}&=& 0 \\
{\overline { v(h_1) v(h_1')}}&=& \delta(h_1-h_1'),
\label{qn2}
\end{eqnarray}
where ${\overline {\left(\ldots\right)}}$ denotes averages over the different realizations of the quenched
velocity field $v(h_1)$. 

We are interested in the following persistence probabilities at a fixed position ${\bf r}$ in space
\begin{eqnarray}
P_1(t, t_0) &=& {\rm {Prob}}\left[h_1({\bf r}, t')\ne h_1({\bf r}, t_0)\,\, {\rm for\,\, all\,\,} t': \,\,
t_0<t'<t_0+t \right]
\label{perh1}\\
P_2(t, t_0) &=& {\rm {Prob}}\left[h_2({\bf r}, t')\ne h_2({\bf r}, t_0)\,\, {\rm for\,\, all\,\,} t': \,\,
t_0<t'<t_0+t \right].
\label{perh2}
\end{eqnarray}
The former represents the probability that the height $h_1({\bf r}, t')$ of the first interface
at a fixed point ${\bf r}$ in space does not return to its value at $t_0$ within the time
interval $[t_0, t_0+t]$. The latter represents the same probability for the second interface.
The persistence probability $P_1(t, t_0)$ for the free Edwards-Wilkinson equation in Eq. (\ref{perh1})
has been studied both analytically and numerically\cite{Krug}. It is known that for $d<2$,
$P_1(t,t_0)$ has a power law decay for large $t$ characterized by a nontrivial
persistence exponent and the value of this exponent depends on whether the waiting time
$t_0=0$ (no return to the initial condition) or $t_0\to \infty$
(no return to a stationary configuration)\cite{Krug}
\begin{eqnarray}
P_1(t, t_0=0) & \sim & t^{-\theta_0^1} \label{p1t0} \\
P_1(t, t_0\to \infty) & \sim & t^{-\theta_s^1} \label{p1ts}
\end{eqnarray}
where the subsripts $0$ and $s$ in the exponents refer respectively to the probabilities of
no return to the initial condition ($t_0=0$) and no return to a stationary configuration ($t_0\to \infty$).
The superscript $1$ refers to the first interface. It turns out that the exponent $\theta_0^1$
is hard to determine analytically and is known only numerically. For example, $\theta_0^1= 1.55\pm 0.02$
in $(1+1)$-dimensions\cite{Krug}. On the other hand, the exponent $\theta_s^1$ was computed
analytically\cite{Krug} by mapping the process $h_1$ in the $t_0\to \infty$ limit to a fBm and then using
a known first-passage result of fBm. For all $d<2$, one gets\cite{Krug}
\begin{equation}
\theta_s^1= (2+d)/4.
\label{thetas1}
\end{equation} 
For $d=1$ this gives $\theta_s^1=3/4$, a result that was recently verified experimentally
in a system of fluctuating $(1+1)$-dimensional steps on Si-Al surfaces\cite{PE}. For $d>2$,
we show that the persistence $P_1(t, t_0\to \infty)$ decays faster than a power law for large $t$,
stretched-exponentially for $2<d<4$ and exponentially for $d>4$. 

In this paper, we study the persistence probability $P_2(t,t_0)$ of the second interface
in Eq. (\ref{h2evol1}). As in the case of the first interface, we find that for large interval $t$
\begin{eqnarray}
P_2(t, t_0=0) & \sim & t^{-\theta_0^2} \label{p2t0} \\
P_2(t, t_0\to \infty) & \sim & t^{-\theta_s^2} \label{p2ts}
\end{eqnarray}
where the superscript $2$ in the exponents refer to the second interface.
While we were not able to compute the exponent
$\theta_0^2$, we calculated the exponent
$\theta_s^2$ analytically for all $d<2$
\begin{equation}
\theta_s^2= (2-d)/8.
\label{thetas2}
\end{equation}
For $d>2$, we argue that the persistence $P_2(t, t_0\to \infty)$ tends to a constant
as $t\to \infty$.

Furthermore, we were able to generalize
this result to the case when the second interface evolves by Eq. (\ref{h2evol1}) but
the first interface is any generic growing surface, evolving not neccesarily  
by the
Edwards-Wilkinson equation. For example the first interface may evolve by the
Kardar-Parisi-Zhang (KPZ) equation\cite{KPZ}. In general, this first interface
will be characterized by a growth exponent $\beta_1$ and a dynamical
exponent $z_1$ defined via the scaling form of the second moment
of the height differences between two points in space,
$\langle \left[h_1({\bf r}_1,{\tau}_1)
- h_1({\bf r}_2,{\tau}_2)\right]^2 \rangle \approx
|\tau_2 - \tau_1|^{2\beta_1} f\left( { {|{\bf r}_1 -
{\bf r}_2|^{z_1}}/{|\tau_2 - \tau_1|}}\right)$.
For example, for the $(1+1)$-dimensional Edwards-Wilkinson equation
one has $\beta_1=1/4$ and $z_1=2$, whereas for the $(1+1)$-dimensional
KPZ equation one has $\beta_1=1/3$ and $z_1=3/2$\cite{HZ}. Our main
result is to show that 
\begin{equation}
\theta_s^2= \beta_1/2.
\label{thetas2g}
\end{equation}    
In particular, Eq. (\ref{thetas2g}) predicts that in $(1+1)$-dimensions, if the
first interface evolves via the KPZ equation, $\theta_s^2=1/6$. 
In Section-VII we show that the numerical results are consistent with
this theoretical prediction.

\section{Limiting Cases}

In this Section we show that in the two limiting cases $d\to 0$ and $d=1$, our model
defined by Eqs. (\ref{h1evol1}) and (\ref{h2evol1}) reduce respectively to
two well studied models.

\subsection{ The Limit $d\to 0$}

In the limit $d\to 0$, there is no `space' variable in the problem. Thus the Laplacian
terms on the right hand side of Eqs. (\ref{h1evol1}) and (\ref{h2evol1}) drop out
and also the noise variables no longer have any ${\bf r}$ dependence. Interpreting
$h_1=x$ and $h_2=y$ as the coordinates $(x,y)$ of a single particle on a $2$-dimensional
plane, the Eqs. (\ref{h1evol1}) and (\ref{h2evol1}) reduce to
\begin{eqnarray}
\frac{dx}{dt}&=& \eta_1(t) \label{mdmx1}\\
\frac{dy}{dt}&=& v\left(x(t)\right) + \eta_2(t) \label{mdmy1}.
\end{eqnarray}
These equations represent precisely the MdM model\cite{MdM} where a single Brownian
particle moves in a $2$-dimensional plane in the presence of a transverse
quenched random velocity field $v(x)$.
This model was originally introduced to study the hydrodynamic dispersion of a
tracer particle in porous rocks\cite{MdM}. While the transport properties in this model
were well understood\cite{MdM1,BoG,Red}, the study of persistence properties
in this model are relatively recent\cite{Redner1,SM}. The persistence probabilities
defined in Eqs. (\ref{perh1}) and (\ref{perh2}) reduce, in this context, to the
following probabilities
\begin{eqnarray}
P_1(t,t_0)&=&{\rm {Prob}}\left[x(t')\ne x(t_0)\,\, {\rm for\,\, all\,\,} t': \,\, t_0<t'<t_0+t \right]
\label{perxmdm}\\
P_2(t,t_0) &=& {\rm {Prob}}\left[y(t')\ne y(t_0)\,\, {\rm for\,\, all\,\,} t': \,\, t_0<t'<t_0+t \right].
\label{perymdm}
\end{eqnarray}
The first probability $P_1(t,t_0)$ is simply the probability of no return to its initial position
of a one dimensional Brownian motion and hence $P_1(t,t_0)\sim t^{-\theta_{0,s}^1}$ where
$\theta_0^1=\theta_s^1=1/2$. The probability $P_2(t,t_0)$ associated with the $y$ coordinate
is nontrivial. Based on heuristic arguments and numerical simulations, Redner showed
that $\theta_0^2=1/4$\cite{Redner1}. More recently, it was proved analytically
that $\theta_0^2=\theta_s^2=1/4$ by mapping the stochastic process $y(t)$ to
a fBm and then using a known first-passage property of the latter\cite{SM}. The result
$\theta_s^2=1/4$ is thus a special case of our result in Eq. (\ref{thetas2}) of the
interface model in the limit $d\to 0$. Incidentally, the exponent $\theta_0^2=1/4$ happens
to be generic for a class of transverse velocity fields 
and occurs even when the tranverse velocity field
is a deterministic but an odd function of $x$, i.e., $v(x)=-v(-x)$\cite{Burkhardt,RK,BG}.

\subsection{ The Case $d=1$}

Let us consider a Rouse polymer chain embedded in a $2$-dimensional plane. The chain consists
of beads connected by harmonic springs\cite{Rouse}. In addition, the chain is advected
by a random layered velocity field as shown in Fig. 1. Let $[x_n(t), y_n(t)]$ denote
the coordinates of the $n$-th bead at time $t$ which evolve
with time according to the following equations of motion
\begin{eqnarray}
\frac{dx_n}{dt} &=& \Gamma\left(x_{n+1}+x_{n-1}-2\,x_n\right) + \eta_1(n,t) \label{evolx1}\\
\frac{dy_n}{dt} &=& \Gamma\left(y_{n+1}+y_{n-1}-2\,y_n\right) + v\left(x_n(t)\right)+ \eta_2(n,t),
\label{evoly1}
\end{eqnarray}
where $\Gamma$ denotes the strength of the harmonic interaction between nearest neighbour beads,
$\eta_1(n,t)$ and $\eta_2(n,t)$ represent the Gaussian thermal noises along the $x$ and $y$
directions respectively and are delta correlated. 
The velocity field $v(x)$ is a random quenched function of $x$ taken to be
a Gaussian with the following 
moments
\begin{eqnarray}
{\overline {v(x)}}&=& 0 \\
{\overline {v(x)v(x')}}&=& \delta(x-x')
\label{qn1}
\end{eqnarray}
For a finite chain with $N$ beads, the Eqs. (\ref{evolx1}) and (\ref{evoly1}) are 
valid only for the $(N-2)$ interior beads. The two boundary beads will have slightly different equations of motion.
However, we will only focus here on an infinitely large chain ($N\to \infty$) so that the system
is translationally invariant along the length of the chain and the boundary conditions are
irrelevant.  
\begin{figure}[htbp]
\epsfxsize=8cm
\centerline{\epsfbox{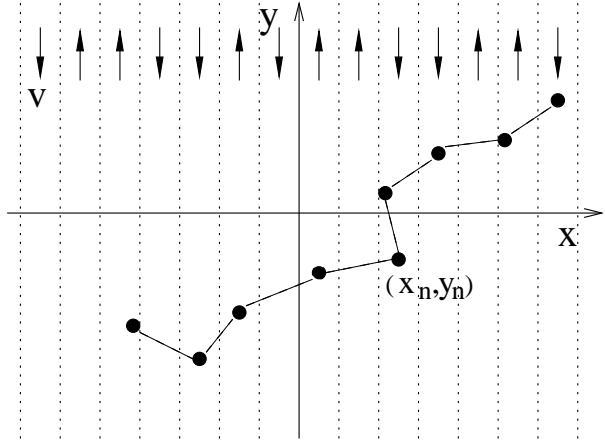}}
\caption{A Rouse chain in a random layered velocity field in $2$ dimensions.} 
\label{fig:chain1}
\end{figure}

Note that in the absence of the harmonic interaction term, i.e., when $\Gamma=0$, this model
reduces precisely to a single particle MdM model discussed in the previous subsection. 
In presence of the harmonic interaction, the
transport properties in this model have been studied recently\cite{OB,WL,JOB}. However the
persistence properties for $\Gamma\ne 0$ have not yet been studied.
One can define the following persistence probabilities,
\begin{eqnarray}
P_1(t,t_0) &=& {\rm {Prob}}\left[x_n(t')\ne x_n(t_0)\,\, {\rm for\,\, all\,\,} t': \,\, t_0<t'<t_0+t \right] 
\label{perx1}\\
P_2(t,t_0) &=& {\rm {Prob}}\left[y_n(t')\ne y_n(t_0)\,\, {\rm for\,\, all\,\,} t': \,\, t_0<t'<t_0+t \right] 
\label{pery1}
\end{eqnarray}
where the former represents the probability that the $x$ coordinate of a bead (say the $n$-th bead)
does not return to its position at time $t_0$ within the time interval $[t_0, t_0+t]$, while the latter
represents the same probability for the $y$ coordinate of the same bead. For an infinite chain, the system is 
translationally invariant along the length of the chain and hence these persistence probabilities
do not depend on the bead label $n$.

Since we are interested in the late time properties, one can conveniently replace the harmonic interaction term in 
Eqs. (\ref{evolx1}) and (\ref{evoly1})
by a continuous Laplacian operator
\begin{eqnarray}
\frac{\partial x}{\partial t}&=& \frac{\partial^2 x}{\partial s^2} + \eta_x(s,t) \label{evolx2}\\
\frac{\partial y}{\partial t}&=& \frac{\partial^2 y}{\partial s^2} + v\left(x\left(s,t\right)\right)
+\eta_y(s,t),
\label{evoly2}
\end{eqnarray}
where $s$ denotes the distance along the chain and we have rescaled the time to
set the coefficient in front of the Laplacian to be unity. Interpreting $x\equiv h_1$
and $y\equiv h_2$, the equations (\ref{evolx2}) and (\ref{evoly2}) reduce precisely
to the $d=1$ version of our model defined in Eqs. (\ref{h1evol1}) and (\ref{h2evol1}).
Thus, substituting $d=1$ in our results for the persistence exponents in Eqs. (\ref{thetas1})
and (\ref{thetas2}), we obtain the persistence exponents for the polymer problem. 
In particular, for the probability of no return to a stationary configuration 
($t_0\to \infty$ limit in Eqs. (\ref{perx1}) and (\ref{pery1})) of the polymer,
we have for large interval $t$,
\begin{eqnarray}
P_{1}(t, t_0\to \infty)& \sim & t^{-3/4} \label{polx1} \\
P_{2}(t, t_0\to \infty)& \sim & t^{-1/8}, \label{poly1} 
\end{eqnarray}
where the exponents $3/4$ and $1/8$ are obtained by substituting $d=1$ in
Eqs. (\ref{thetas1}) and (\ref{thetas2}) respectively.
Interestingly, for $\Gamma=0$, i.e., in the single particle MdM model,
the corresponding probabilities decay as, $P_1(t, t_0\to \infty)\sim t^{-1/2}$
and $P_2(t, t_0\to \infty) \sim t^{-1/4}$, as discussed in the previous
subsection. Thus, switching on the harmonic interaction strength $\Gamma$ has
opposite effects on $P_1$ and $P_2$. While the probability $P_1(t, t_0\to \infty)$
decays faster as $\sim t^{-3/4}$ (compared to $\sim t^{-1/2}$ when $\Gamma=0$) 
in presence of interaction $\Gamma$, 
the probability
$P_2(t, t_0\to \infty)$ decays slower as $\sim t^{-1/8}$ when $\Gamma$ is switched on
(compared to $\sim t^{-1/4}$ when $\Gamma=0$).

\section{The Strategy to Compute the Persistence Exponents}

The stochastic process $h_1({\bf r}, t)$ in Eq. (\ref{h1evol1}) for a fixed ${\bf r}$, 
as a function of time $t$, is Gaussian albeit non-Markovian. The non-Markovian property 
arises due to the Laplacian term that generates interaction between heights at two
different space points. The process $h_2({\bf r}, t)$ in Eq. (\ref{h2evol1}) for fixed ${\bf r}$
is similarly non-Markovian and moreover it is non-Gaussian due to the quenched velocity
field in Eq. (\ref{h2evol1}). Analytical calculation of the persistence exponent
is known to be very hard for a non-Markovian process even if the process is 
Gaussian\cite{Review}. For a non-Gaussian process it is even harder in general.
However, in certain cases, it may be possible to map the relevant stochastic
process into a fBm and then one can use a known first-passage property of fBm
to calculate the persistence exponent analytically. This strategy has been successful
in the past to compute analytically the persistence exponents in a number of problems
even though the relevant processes were non-Gaussian and/or non-Markovian\cite{SM,Krug,Krug1,MB1,M3,Krug2}. 
In particular, this strategy was used in Ref. \cite{Krug} to calculate the exponent
$\theta_s^1=(2+d)/4$ in Eq. (\ref{thetas1}) for the first interface evolving
freely with a Edwards-Wilkinson dynamics. 
Here we exploit the same strategy to compute the exponent
$\theta_s^2=(2-d)/8$ in Eq. (\ref{thetas2}) for the second interface.
 
Before proceeding further, it is then useful at this point to summarize the definition and the known 
first-passage property of a fBm. A stochastic process $X(t)$ (with zero mean $E[X(t)]=0$) is called
a fBm if its incremental two-time correlation function $C(t_1,t_2)= 
E\left[\left(X(t_1)-X(t_2)\right)^2\right]$ is (i) {\em stationary}, i.e., depends only
on the time difference $|t_2-t_1|$ and (ii) grows asymptotically as a power law
\begin{equation}
C(t_1,t_2)\sim |t_2-t_1|^{2H}, \quad\quad |t_2-t_1|>>1.
\label{fbm1}
\end{equation}
The parameter $0<H<1$ is called the Hurst exponent that chracterizes the fBm\cite{MN} and
$E[\dots]$ denotes the expectation value over all realizations of the process $X(t)$. For example,
an ordinary Brownian motion corresponds to a fBm with $H=1/2$. The zero crossing properties
of a fBm has been studied extensively in the past\cite{Berman,HEM,DY,Krug}. The particular
property that is useful for our purpose is the fact that the probability that a fBm does not cross
zero upto time $t$ decays as a power law at late times, $P(t)\sim t^{-\theta}$ with 
$\theta=1-H$\cite{DY,Krug}. Note that this result $\theta=1-H$ holds for any zero mean process that
satisfies the defining characteristics (i) and (ii) above of a fBm. In particular,
it holds even if the process is non-Gaussian and/or non-Markovian
as long it satisfies (i) and (ii) above.  A simple proof of this 
result is 
given in Ref. \cite{Krug}. Our strategy in the next Section would be to first show that
indeed for $d<2$, the processes $h_1({\bf r}, t)$ and $h_2({\bf r}, t)$ 
in Eqs. (\ref{h1evol1}) and (\ref{h2evol1}), for fixed ${\bf r}$ and in the limit 
$t_0\to \infty$, do satisfy
the characteristics (i) and (ii) of a fBm with Hurst exponents $H_1=(2-d)/4$ and $H_2=(6+d)/8$
respectively and then use the powerful result $\theta=1-H$ stated above to prove that
$\theta_s^1=(2+d)/4$ and $\theta_s^2=(2-d)/8$. For the free interface $h_1$ in Eq. (\ref{h1evol1})
this mapping to the fBm was already done in Ref. \cite{Krug}, but we will include it Section 
V-A for completeness.
The corresponding calculations for $h_2$ are new and will be detailed in Section V-B.  

\section{Two-time Correlation Functions}

In order to make use of the first-passage property of fBm mentioned in the Section-IV, we need
to calculate the incremental correlation functions,
\begin{eqnarray}
C_1(t_1,t_2,t_0)&=& E\left[ {\left( h_1({\bf r}, t_0+t_1)-h_1({\bf r}, t_0+t_2)\right)}^2\right]\label{c1in1}\\
C_2(t_1,t_2,t_0)&=& E\left[ {\left( h_2({\bf r}, t_0+t_1)-h_2({\bf r}, t_0+t_2)\right)}^2\right]\label{c2in1}
\end{eqnarray}
where $E\left[\dots\right]$ denotes an average over both thermal noises as well as the disorder.
Note that while for the first interface $h_1$, the expectation $E\left[\dots\right]$ is only
over the thermal noise, for the second interface it includes averages over both the thermal
noise as well as the disorder. In Section V-A, we calculate $C_1(t_1,t_2,t_0)$ which
is rather straightforward. The computation of $C_2(t_1,t_2,t_0)$ is more involved
and is detailed in Section V-B.

\subsection{Incremental correlation function for the first interface}

Since the Eq. (\ref{h1evol1}) is linear one can solve it by the standard Fourier transform technique.
We define the Fourier transform, ${\tilde h}_1({\bf k},t) = \int d^d{\bf r}\, h_1({\bf r},t)
\exp(i {\bf k}.{\bf r})$. Taking the Fourier transform of Eq. (\ref{h1evol1}) one gets
\begin{equation}
{\tilde h}_1({\bf k},t) = e^{-k^2 t} \int^{t}_{0} e^{k^2 t^{\prime}}
 \tilde{\eta}_1 ({\bf k},t^{\prime}) dt^{\prime},
\label{h1four}
\end{equation} 
where we have assumed that the system starts from a flat initial condition $h_1({\bf r}, t=0)=0$.
Using the noise correlator in Eq. (\ref{eta1}) one can then easily show that
\begin{equation}
\langle {\tilde h}_1 ({\bf k},t_1) {\tilde h}_1 (-{\bf k},t_2) \rangle
= {1 \over 2k^2}[e^{-k^2|t_2 - t_1|} - e^{-k^2(t_1 + t_2)}] e^{-k^2 a^2}.
\label{corrh1four}
\end{equation}
The auto-correlation function in real space can then be obtained as
\begin{eqnarray}
A_1(t_1,t_2) &=& \langle h_1({\bf r},t_1) h_1({\bf r},t_2) \rangle \nonumber \\
&=& \int {d^d{\bf k} \over (2\pi)^d} {e^{-k^2 a^2} \over {2 k^2}}
\left[-(1 - e^{-k^2 |t_2 - t_1|}) + (1 - e^{-k^2 (t_1 + t_2)})\right].
\label{auto1}
\end{eqnarray}
The right hand side of Eq. (\ref{auto1}) is a sum of two integrals each of which is convergent
and can be easily done in closed form and one gets\cite{Krug}
\begin{equation}
A_1(t_1,t_2) = {{a_0} \over (2-d)} \left[(t_1+t_2+a^2)^{1-{d \over 2}}
- (|t_2 - t_1| + a^2)^{1 - {d \over 2}}\right]
\label{A1}
\end{equation}
where $a_0=(4\pi)^{-d/2}$. Thus, as $t\to \infty$, $A_1(t,t)=\langle h_1^2({\bf r},t)\rangle\sim 
t^{1-d/2}=t^{2\beta_1}$ for $d<2$. One the other hand, for $d>2$, $A_1(t,t)\to a_0/((d-2) a^{d-2})$
as $t\to \infty$. Thus while for $d>2$ the surface becomes smooth in the stationary state, for $d<2$ 
the surface roughens with time and the fluctuations grow as a power law in an infinite system
\begin{equation}
\langle h_1^2({\bf r},t)\rangle\sim t^{2\beta_1},\quad\quad {\rm where}\,\,\, \beta_1=(2-d)/4.
\label{beta1}
\end{equation} 

The incremental two-time correlation function defined in Eq. (\ref{c1in1}) can then be writen as
\begin{eqnarray}
C_1(t_1, t_2, t_0) &=& \langle {\left[h_1({\bf r},t_1+t_0) - h_1({\bf r},t_2+t_0)\right]}^2
\rangle \nonumber \\
&=& A_1(t_1+t_0,t_1+t_0) + A_1(t_2+t_0,t_2+t_0) - 2 A_1(t_1+t_0,t_2+t_0).
\label{corr1}
\end{eqnarray}
Using Eq. (\ref{A1}) and taking the limit $t_0\to \infty$ one finds that for $d<2$
\begin{eqnarray}
C_1(t_1,t_2,t_0\to \infty) &=& {{a_0} \over (2-d)} \left[2 (|t_2 - t_1| + a^2)^{1 -
{d \over 2}} - 2 a^{2-d}\right] \nonumber \\
&\sim & |t_2 - t_1|^{1 - {d \over 2}} \quad \quad {\rm for} \,\,\, |t_2-t_1|>>a^2.
\label{fbmh1}
\end{eqnarray} 
Comparing with the defining property of the fBm in Eq. (\ref{fbm1}) we see that for
$d<2$ and in the limit 
$t_0\to \infty$, i.e., when one observes the stochastic process $h_1({\bf r}, t)$ at a fixed 
point ${\bf r}$ in space as a function of time after a waiting time $t_0\to \infty$, the
process is a fBm with a Hurst exponent $H_1=(2-d)/4$. This then proves that
the persistence $P_1(t, t_0\to \infty)\sim t^{-\theta_s^1}$ as $t\to \infty$ where
\begin{equation}
\theta_s^1= 1-H_1= (2+d)/4.
\label{thetas1p}
\end{equation}
In the limit $d\to 0$, Eq. (\ref{thetas1p}) gives $\theta_s^1=1/2$ (the classical Brownian motion result)
and for $d=1$, one gets $\theta_s^1=3/4$, a result that was first predicted in Ref.\cite{Krug}
and was subsequently verified experimentally\cite{PE}.

We now turn to the persistence probability $P_1(t,t_0\to \infty)$ for $d>2$. For $d>2$,
the physics of the process is rather different. The fluctutaions of height $h_1$ no longer grow
at late times, but rather saturates to a constant, $\langle h_1^2({\bf r},t)\rangle=A_1(t,t)\to
a_0/((d-2) a^{d-2})$ as $t\to \infty$. Thus, for $d>2$, the cut-off $a>0$ is essential. 
Besides, the relevant process $h_1({\bf r}, t)$, at a fixed ${\bf r}$ but as a function of time, 
no longer satisfies the properties
of a fBm. Thus one can no longer use the first-passage property of fBm to
compute the persistence $P_1(t, t_0\to \infty)$. To make progress, we note that
for $d>2$, it follows from Eq. (\ref{A1}) that in the limit when both $t_1$ and $t_2$
are large but their difference $|t_2-t_1|=t$ kept fixed, the 
auto-correlation
function becomes stationary, i.e, only a function of $t=|t_2-t_1|$ and decays as a power law
\begin{equation}
A_1(t_1,t_2)\approx \frac{a_0}{(d-2)}\, \frac{1}{\left(|t_2-t_1|+a^2\right)^{d/2-1}}.
\label{autos1}
\end{equation}
Thus, in this limit, $h_1$ is a Gaussian stationary process with an auto-correlator
$A_1(|t_2-t_1|=t)\sim t^{-(d/2-1)}$. Interestingly, exactly the same
Gaussian stationary process also represents the total magnetization
of a manifold in nonequilibrium critical dynamics within the mean field theory\cite{MB2}.
In general, the calculation of persistence of a Gaussian
stationary process with an algebraically decaying correlator is nontrivial.
However, it was pointed out in Ref. \cite{MB2} that one can use a
powerful theorem due to Newell and Rosenblatt\cite{NR} to obtain
useful bounds on the persistence property of such a Gaussian process.
The Newell-Rosenblatt theorem states that if the stationary auto-correlator
of a Gaussian process decays algebraically as $A(t)\sim t^{-\alpha}$ for large
time difference $|t_2-t_1|=t$, then the persistence P(t) (i.,e., the probability
of no zero crossing over a time interval $t$) has the following asymptotic behaviors,
\begin{eqnarray}
& P(t) & \sim  \exp[-K_1 t] \quad\quad {\rm for} \,\,\, \alpha>1 \nonumber \\
\exp[-K_2 t^{\alpha} \ln t] \le & P(t)& \le \exp[-K_3 t^{\alpha}] \quad\quad {\rm for}\,\,\, 0<\alpha<1,
\label{nrth}
\end{eqnarray}
where $K_1$, $K_2$ and $K_3$ are some constants. In the borderline case $\alpha=1$, one has
additional logarithmic correction. Applying this theorem to our interface problem upon
identifying $\alpha=d/2-1>0$, we conclude that for large $t$  
\begin{eqnarray}
&P_1(t, t_0\to \infty)&\sim  \exp[-K_1 t] \quad\quad {\rm for}\,\,\, d>4 \nonumber \\
\exp[-K_2 t^{d/2-1} \ln t] \le & P_1(t, t_0\to \infty)& \le \exp[-K_3 t^{d/2-1}] \quad\quad {\rm for}\,\,\, 2<d<4.
\label{nrth1}
\end{eqnarray}
Thus the persistence decays exponentially at late times for $d>4$ and stretched-exponentially
for $2<d<4$. This is to be contrasted with the power law decay $P_1(t, t_0\to \infty)\sim t^{-\theta_s^1}$
for $d<2$. 

\subsection{Incremental correlation function for the second interface} 

In this subsection we show that in the limit $t_0\to \infty$ and for $d<2$, even $h_2$ is a fBm
process and one can calculate exactly the corresponding Hurst exponent $H_2$
and hence the persistence exponent $\theta_s^2=1-H_2$. Taking the Fourier transform of
Eq. (\ref{h2evol1}) one gets
\begin{equation}
{\tilde h}_2({\bf k},t) = e^{-k^2 t} \int^{t}_{0} dt^{\prime}
e^{k^2 t^{\prime}} \left[\tilde{\eta}_2 ({\bf k},t^{\prime}) + \int e^{i
{\bf k}.{\bf r}^{\prime}} v\left(h_1({\bf r}^{\prime},{t}^{\prime})\right)
d^d{\bf r}^{\prime}\right].
\label{fth2}
\end{equation}
Alternatively in real space, one can write
\begin{equation}
h_2({\bf r},t) = \int^{t}_{0} dt^{\prime} \int
d^d{\bf r}^{\prime} {e^{-({\bf r} - {\bf r}^{\prime})^2/{4(t-t^{\prime})}}
\over {[4 \pi (t-t^{\prime})]^{d/2}}} \left[\eta_2({\bf r}^{\prime},t^{\prime})
 + v\left(h_1({\bf r}^{\prime},t^{\prime})\right)\right].
\label{rsh2}
\end{equation}

Then the auto-correlation function, averaged over both the thermal noise and the disorder, is given by
\begin{equation}
A_2(t_1,t_2) = E\left[h_2({\bf r},t_1)h_2({\bf r},t_2)\right]={\overline {\langle h_2({\bf r},t_1) h_2({\bf 
r},t_2) \rangle}}
= I_1(t_1,t_2) + I_2(t_1,t_2),
\label{A2}
\end{equation}
where $I_1(t_1,t_2) $ counts the contribution due to the noise-noise correlator $\langle \eta_2\eta_2 \rangle$
and has the same expression as the auto-correlation function of the first interface,
\begin{equation}
I_1(t_1,t_2)= A_1(t_1,t_2)={{a_0} \over (2-d)} \left[(t_1+t_2+a^2)^{1-{d \over 2}}
- (|t_2 - t_1| + a^2)^{1 - {d \over 2}}\right]
\label{I1}
\end{equation}
On the other hand $I_2(t_1,t_2)$ counts the
contribution due to the random velocity term and is given by
\begin{equation}
I_2(t_1,t_2) =  \int^{t_1}_{0} d{\tau}_1 \int^{t_2}_{0} d{\tau}_2
\int d^d{\bf r}^{\prime}_1
 {e^{-({\bf r} - {\bf r}^{\prime}_1)^2/{4(t_1-{\tau}_1)}}
\over {[4 \pi (t_1-{\tau}_1)]^{d/2}}}
\int  d^d{\bf r}^{\prime}_2
 {e^{-({\bf r} - {\bf r}^{\prime}_2)^2/{4(t_2-{\tau}_2)}}
\over {[4 \pi (t_2-{\tau}_2)]^{d/2}}}~~{\overline {\langle v.v \rangle}}
\label{I2}
\end{equation}
where
\begin{eqnarray}
{\overline {\langle v.v \rangle}} &=&
{\overline {\langle v[h_1({\bf r}^{\prime}_1,{\tau}_1)]
v[h_1({\bf r}^{\prime}_2,{\tau}_2)] \rangle}} \nonumber \\
&=& {\langle {\delta(h_1({\bf r}^{\prime}_1,{\tau}_1) -
h_1({\bf r}^{\prime}_2,{\tau}_2))} \rangle} \nonumber \\
&=& \int_{-\infty}^{\infty} \frac{dq}{2\pi}\, \langle e^{i\,q\,[h_1({\bf r}^{\prime}_1,{\tau}_1) -
h_1({\bf r}^{\prime}_2,{\tau}_2)]} \rangle.
\label{vvcorr}
\end{eqnarray}
In Eq. (\ref{vvcorr}) we have first performed the disorder average 
which gives the expression on the second line. Next we used
an integral representation of the delta function in the third line.
The thermal
average over the delta function can be done by noting that for a Gaussian
process $h$, $\langle \delta(h) \rangle = \int^{\infty}_{-\infty} {dq \over
(2\pi)} \langle e^{iqh} \rangle = \int^{\infty}_{-\infty} {dq \over (2\pi)}
e^{-q^2\langle h^2 \rangle/2} = 1/\sqrt{2\pi \langle h^2 \rangle}$; in
our case $h_1$ being a Gaussian process, $(h_1({\bf r}^{\prime}_1,{\tau}_1) -
h_1({\bf r}^{\prime}_2,{\tau}_2))$ is also Gaussian. This gives
\begin{equation} 
{\overline {\langle v.v \rangle}}=
{1 \over {\sqrt {2\pi \langle (h_1({\bf r}^{\prime}_1,{\tau}_1) -
h_1({\bf r}^{\prime}_2,{\tau}_2))^2 \rangle}}}.
\label{vvcorr1}
\end{equation}

Using Eq. (\ref{corrh1four})
one can easily compute the following two-time correlation function
\begin{eqnarray}
\langle h_1({\bf r}^{\prime}_1,{\tau}_1) h_1({\bf r}^{\prime}_2,{\tau}_2)
\rangle &=& \int \frac{d^d {\bf k}}{(2\pi)^d}\, \langle {\tilde h}_1({\bf k},t_1){\tilde h}_1
(-{\bf k},t_2)\rangle \,
e^{i{\bf k}.({\bf r}^{\prime}_1-{\bf r}^{\prime}_2)} \nonumber \\
&=& {1 \over 2} \int^{\tau_1 + \tau_2 + a^2}_{|\tau_1 - \tau_2| + a^2}
d\tau  {e^{-({\bf r}^{\prime}_1 - {\bf r}^{\prime}_2)^2/4\tau }
\over {(4\pi \tau )}^{d/2}}.
\label{h1tt}
\end{eqnarray}
For $d<2$, one can safely set the cut-off $a\to 0$ for simplicity and using Eq. (\ref{h1tt})
one gets
\begin{equation}
\langle \left[h_1({\bf r}^{\prime}_1,{\tau}_1)
- h_1({\bf r}^{\prime}_2,{\tau}_2)\right]^2 \rangle =
\frac{a_0}{(2-d)}\left[ (2\tau_1)^{1 - {d \over 2}}
+ (2\tau_2)^{1 - {d \over 2}} - (2-d) \int^{\tau_1 +
\tau_2}_{|\tau_1 - \tau_2|}
d\tau\, {\tau}^{-d/2}\, e^{-({\bf r}^{\prime}_1 - {\bf r}^{\prime}_2)^2/4\tau }\right].
\label{M}
\end{equation}
Eqs. (\ref{M}) and (\ref{vvcorr1}) thus fully specify $I_2(t_1,t_2)$ in Eq. \ref{I2} which can be further 
simplified
by making the change of variables: ${\bf r}^{\prime}_1 - {\bf r} = {\bf u}$ and
${\bf r}^{\prime}_2 - {\bf r}^{\prime}_1 ={\bf z}$.
The integral over ${\bf u}$ is a Gaussian that can be easily performed.
After a few steps of elementary algebra we get
\begin{eqnarray}
I_2(t_1,t_2)& = & \sqrt{2(2-d)}\, (4\pi)^{(d-2)/4}
\int d^d{\bf z}\, 
\frac{e^{-{{\bf z}^2}/{4(t_1+t_2-\tau_1-\tau_2)}}}
{ {\left[4\pi\left(t_1+t_2-\tau_1-\tau_2\right)\right]}^{d/2} }
\times \nonumber \\
& &\int^{t_1}_0 d{\tau_1} \int^{t_2}_0 d{\tau_2}
{\left[\left(2\tau_1\right)^{1-{d \over 2}} + \left(2\tau_2\right)^{1-{d \over 2}}
- (2-d)  \int^{\tau_1 + \tau_2}_{|\tau_1 - \tau_2|}
d\tau \, {\tau}^{-d/2}\, 
e^{-{{\bf z}^2}/{4\tau}}\right]}^{-1/2}.
\label{I2simple} 
\end{eqnarray}
Note that when $t_1=t_2=t$ a simple power counting in Eq. (\ref{I2simple})
shows that $I_2(t,t)\sim t^{(d+6)/4}$ for large $t$. On the other hand, it follows
from Eq. (\ref{I1}) that $I_1(t,t)\sim t^{(2-d)/2}$ for large $t$. Thus $I_2(t,t)$
grows faster than $I_1(t,t)$ for large $t$. Thus it follows from
Eq. (\ref{A2}) that
\begin{equation}
E\left[h_2^2({\bf r}, t)\right]=A_2(t,t)\sim t^{2\beta_2}, \quad\quad {\rm where}\,\,\,\beta_2=(6+d)/8.
\label{beta2}
\end{equation}
In particular for $d=1$, the result $\beta_2=7/8$ agrees with that of Ref.\cite{JOB} derived
in the context of a Rouse chain in a random tranverse velocity field.

The incremental correlation function defined in Eq. (\ref{c2in1}) can then be written in terms of
the auto-correlation function
\begin{eqnarray}
C_2(t_1, t_2, t_0) &=& E\left[{\left(h_2({\bf r},t_1+t_0) - h_2({\bf r},t_2+t_0)\right)}^2\right]
\rangle \nonumber \\
&=& A_2(t_1+t_0,t_1+t_0) + A_2(t_2+t_0,t_2+t_0) - 2 A_2(t_1+t_0,t_2+t_0) \nonumber \\
&=& C_1(t_1,t_2,t_0) +C^{I_2}_2,
\label{c2in2}
\end{eqnarray}
where in going from the second to the third line in Eq. (\ref{c2in2}) we have used 
Eq. (\ref{A2}), the fact that $C_1(t_1,t_2,t_0)= 
I_1(t_1+t_0,t_1+t_0)+I_1(t_2+t_0,t_2+t_0)-2I_1(t_1+t_0,t_2+t_0)$ and the contribution
$C^{I_2}_2$ coming from the $I_2$ terms is defined as
\begin{equation}
C^{I_2}_2 = I_2(t_1+t_0,t_1+t_0) + I_2(t_2+t_0,t_2+t_0) -2\, I_2(t_1+t_0,t_2+t_0).
\label{C2I2}
\end{equation}

A close inspection of Eq. (\ref{I2simple}) reveals that for large $t_1$ and $t_2$ the dominant contribution
to the integral comes from the region where $(t_1 + t_2 - \tau_1 - \tau_2) \to 0$. To capture
the leading behavior of $I_2(t_1,t_2)$ for large $t_1$ and $t_2$ it is then sufficient
to make the replacement $e^{-{{\bf z}^2}/{4(t_1+t_2-\tau_1-\tau_2)}}/
{\left[4\pi (t_1+t_2-\tau_1-\tau_2)\right]}^{d/2} \to \delta({\bf z})$ in Eq. (\ref{I2simple}). Thus to 
leading order
for large $t_1$ and $t_2$ one gets
\begin{equation}
I_2(t_1,t_2)\approx \sqrt{2(2-d)}\, (4\pi)^{(d-2)/4}
\int_0^{t_1}d\tau_1 \int_0^{t_2} d\tau_2 {\left[(2\tau_1)^{1-d/2}+(2\tau_2)^{1-d/2}-2(\tau_1+\tau_2)^{1-d/2}
+ 2|\tau_1-\tau_2|^{1-d/2}\right]}^{-1/2}.
\label{I2approx}
\end{equation}
For large $t_0$, all the arguments of $I_2$ on the r.h.s of Eq. (\ref{C2I2}) are large even
though $t_1$ and $t_2$ may not be large. Hence one can use the asymptotic
expression of $I_2$ in Eq. (\ref{I2approx}). 
Substituting Eq. (\ref{I2approx}) in Eq. (\ref{C2I2}) and rearranging the
domains of integration one finds for large $t_0$ 
\begin{equation}
C^{I_2}_2 \approx \sqrt{2(2-d)}\, (4\pi)^{(d-2)/4}
\int^{t_2+t_0}_{t_1+t_0} d{\tau_1} \int^{t_2+t_0}_{t_1+t_0} d{\tau_2}
{\left[(2\tau_1)^{1-d/2}+(2\tau_2)^{1-d/2}-2(\tau_1+\tau_2)^{1-d/2}
+ 2|\tau_1-\tau_2|^{1-d/2}\right]}^{-1/2}.
\label{C2I2ap}
\end{equation}
Making a further change of variables $x_1=\tau_1-(t_1+t_0)$ and $x_2=\tau_2-(t_2+t_0)$ 
one finds that to leading order for large $t_0$ the r.h.s of Eq. (\ref{C2I2ap}) becomes
independent of $t_0$ and depends only on $|t_2-t_1|$,
\begin{equation}
C^{I_2}_2 \approx \sqrt{(2-d)}\, (4\pi)^{(d-2)/4}\int_0^{|t_2-t_1|}dx_1\int_0^{|t_2-t_1|}dx_2 
|x_1-x_2|^{(d-2)/4}.
\label{C2I2fi}
\end{equation}
Performing the double integral in Eq. (\ref{C2I2fi}) one finally gets
\begin{equation}
C^{I_2}_2 \approx \frac{32\, \sqrt{(2-d)}}{(d+2)(d+6)}\, (4\pi)^{(d-2)/4}\,\, 
|t_2-t_1|^{(6+d)/4}.
\label{C2I2fi1}
\end{equation}
Substituting the results from Eqs. (\ref{fbmh1}) and (\ref{C2I2fi1}) on the r.h.s of
Eq. (\ref{c2in2}) one gets for large $t_0$
\begin{equation}
C_2(t_1,t_2, t_0\to \infty)\approx b_1 |t_2-t_1|^{1-d/2} + b_2 |t_2-t_1|^{(6+d)/4},
\label{fbmh21}
\end{equation}
where $b_1$ and $b_2$ are two constants that can be read off Eqs. (\ref{fbmh1}) and (\ref{C2I2fi1})
respectively. Since the exponent $(6+d)/4> (1-d/2)$ one gets for large $|t_2-t_1|$,
\begin{equation}
C_2(t_1,t_2, t_0\to \infty)\sim |t_2-t_1|^{(6+d)/4}.
\label{fbmh2}
\end{equation}
Comparing with Eq. (\ref{fbm1}) one thus finds that for large $t_0$ and $d<2$, the process $h_2$
is also a fBm with
\begin{eqnarray}
H_2 &=& (6+d)/8 \\
\theta_s^2 &=& 1 - H_2 = (2-d)/8 = \beta_1/2.
\label{thetas2f}
\end{eqnarray}
In the limit $d\to 0$, one thus recovers the result $\theta_s^2=1/4$ for a single particle MdM
model\cite{Redner1,SM}. For $d=1$, we get $\theta_s^2=1/8$, the exact persistence exponent
for the Rouse chain in a transverse velocity field.  

The above results are valid for $d<2$. For $d>2$, one needs to keep the cut-off $a$ finite.
Carrying out a similar analysis as in the $d<2$ case but keeping $a$ finite, one can show that in the limit
$t_0\to \infty$, the incremental correlation function has the following asymptotic behavior for all
$d>2$,
\begin{equation}
C_2(t_1, t_2, t_0\to \infty)\sim |t_2-t_1|^2, \quad\quad {\rm for}\,\,\, |t_2-t_1|>> a^2.
\label{dgt4}
\end{equation}
Thus for $d>2$, $h_2$ is a fBm with $H_2=1$ which indicates that $\theta_s^2=0$ for all
$d>2$. This means that the persistence $P_2(t, t_0\to \infty)$ tends to a constant
at large time $t$ for all $d>2$. Indeed, Eq. (\ref{dgt4}) indicates that the surface
$h_2$ grows ballistically at late times and with a finite probability $h_2$ does not
return to its starting position at $t_0$ over the time interval $[t_0, t+t_0]$. 
In combination with the result in Eq. (\ref{thetas2f}), one thus concludes
that the exponent $\theta_s^2=(2-d)/8$ for $d<2$ tends to $0$ as $d\to 2^{-}$
and then sticks to $\theta_s^2=0$ for all $d>2$.
Note that this behavior of $P_2(t, t_0\to \infty)$ for $d>2$ is quite
opposite to the corresponding $d>2$ behavior of the first interface $h_1$ for which
$P_1(t, t_0\to \infty)$ decays
faster than a power law at large $t$ as was shown in the previous subsection.

\section{Generalization to other Growing Interfaces}

In this Section, we consider a generalization of our model of coupled interfaces. In this generalized version, 
while the second interface height $h_2$ still evolves via Eq. (\ref{h2evol1}), the first interface
height $h_1$ may correspond to any generically growing interface, not necessarily evolving via
the Edwards-Wilkinson equation (\ref{h1evol1}). For example, $h_1$ may evolve via the KPZ equation\cite{KPZ}
\begin{equation}
{{\partial h_1} \over {\partial t}} = {\nabla}^2 h_1 + \lambda ({\nabla h_1})^2
+ \eta_1({\bf r},t).
\label{h1kpz} 
\end{equation}
In general, we will consider a generically growing interface $h_1$ characterized by the following
dynamical scaling of its space-time correlation function\cite{HZ}
\begin{equation}
\langle [h_1({\bf r}^{\prime}_1,{\tau}_1)
- h_1({\bf r}^{\prime}_2,{\tau}_2)]^2 \rangle \approx
|\tau_2 - \tau_1|^{2\beta_1} f\left({{|{\bf r}^{\prime}_1 -
{\bf r}^{\prime}_2|^{z_1}} \over {|\tau_2 - \tau_1|}}\right)
\label{scaling1}
\end{equation}
where $\beta_1>0$ is the growth exponent, $z_1$ is the dynamical exponent and $f(y)$ is the
dynamical scaling function which approaches a constant as $y\to 0$, $f(0)=C$ and decays
for large $y$. For example,
for the $(1+1)$-dimensional Edwards-Wilkinson equation, $\beta_1=1/4$ and $z_1=2$ whereas
for the $(1+1)$-dimensional KPZ equation, $\beta_1=1/3$ and $z_1=3/2$\cite{HZ}.
Note that for a $(1+1)$-dimensional KPZ equation, one looses the symmetry $h\to -h$.
Hence, the persistence exponent $\theta_s^1$ associated with $h_1$ will
be different depending on whether the process $h_1$ stays above its mean value
or below its mean value\cite{Krug1,M3}. In this paper, we focus only on
the second interface $h_2$ for which there is only one persistence exponent
$\theta_s^2$. The goal of this section is to show that quite generically
$\theta_s^2=\beta_1/2$. 

We follow a similar route as in Section V-B and start with the calculation of
the autocorrelation function of the $h_2$. All the steps from Eq. (\ref{fth2})
to Eq. (\ref{vvcorr}) remain unchanged since one doesn't use any information
about $h_1$ till Eq. (\ref{vvcorr}).  One uses the specific information about
$h_1$ for the first time in evaluating the average ${\overline {\langle v.v \rangle}}$
in Eq. (\ref{vvcorr1}). In Section V-B, the process $h_1$ is Gaussian at all times since
it evolves via the linear Edwards-Wilkinson equation (\ref{h1evol1}). This fact
that $h_1$ is Gaussian was used explicitly in evaluating the thermal average
$\langle \exp \left[i\,q\,[h_1({\bf r}^{\prime}_1,{\tau}_1) -
h_1({\bf r}^{\prime}_2,{\tau}_2)\right]\rangle$ in Eq. (\ref{vvcorr}) which
led to the result in Eq. (\ref{vvcorr1}).
For a generic non-Gaussian interface $h_1$ one can use this step to evaluate the
thermal average in Eq. (\ref{vvcorr}). To make progress, let us first denote
$h= \left[h_1({\bf r}^{\prime}_1,{\tau}_1) -
h_1({\bf r}^{\prime}_2,{\tau}_2)\right]$. Then Eq. (\ref{vvcorr}) gives
\begin{eqnarray}
{\overline {\langle v.v \rangle}} &=& \int_{-\infty}^{\infty} \frac{dq}{2\pi}\, \langle e^{i\,q\,h}\rangle 
\nonumber \\
&=& \int_{-\infty}^{\infty} \frac{dq}{2\pi}\, \int dh\, e^{i\,q\,h}\, P(h)
\label{vvcorr2}
\end{eqnarray}  
where $P(h)$ is the normalized probability distribution of the variable $h$. For a generic
interface, one expects the normalized distribution to have the scaling form, $P(h)={1\over {\sqrt{\langle 
h^2\rangle}}}\,F\left({h\over {\sqrt{\langle h^2\rangle}}}\right)$. Substituting this scaling form in
Eq. (\ref{vvcorr2}) and rescaling, one finds the following scaling
\begin{equation}
{\overline {\langle v.v \rangle}} \sim {1\over {\sqrt{\langle h^2\rangle}}}= {1 \over {\sqrt { \langle 
(h_1({\bf r}^{\prime}_1,{\tau}_1) -
h_1({\bf r}^{\prime}_2,{\tau}_2))^2 \rangle}}}.
\label{vvcorr3}
\end{equation}
Next we substitute the generic dynamical scaling form in Eq. (\ref{scaling1}) for the correlation
function in the denominator of Eq. (\ref{vvcorr3}) and use the resulting expression of
${\overline {\langle v.v \rangle}}$ on the r.h.s of Eq. (\ref{I2}). The subsequent evaluation
of the integral $I_2(t_1,t_2)$ for large $t_1$ and $t_2$ followed by the evaluation
of the incremental correlation function $C_2(t_1,t_2,t_0)$ can be done by following
an identical analysis as in Section V-B which we do not repeat here. 
After performing these steps one finds that for large $t_0\to \infty$
\begin{equation}
C_2(t_1,t_2, t_0\to \infty) \sim b_1 |t_2-t_2|^{1-d/2} + c_1 |t_2-t_1|^{2-\beta_1}, 
\label{cingen1}
\end{equation}
where $b_1$ and $c_1$ are unimportant constants. Thus, $h_2$ satisfies the defining 
property in Eq. (\ref{fbm1}) of a fBm with a Hurst exponent given by
\begin{equation}
H_2= {\rm max}\left[(2-d)/4, 1-\beta_1/2\right].
\label{hurstg1}
\end{equation}
In particular, for $\beta_1< 1+d/2$ (which seems to be the case for most interfaces), one gets 
$H_2=1-\beta_1/2$. This then 
leads to the persistence exponent
\begin{equation}
\theta_s^2=1-H_2= \beta_1/2 .
\label{hurstg2}
\end{equation}
For example, for a $(1+1)$-d KPZ interface $h_1$ for which $\beta_1=1/3$, one gets
\begin{equation}
H_2= 1-\beta_1/2=5/6; \quad\quad\quad \theta_s^2=1/6.
\label{kpze}
\end{equation}

\section{Numerical Simulations}

\begin{figure}
\begin{center}
\epsfxsize=9.0cm \epsfysize=8.0cm
\epsfbox{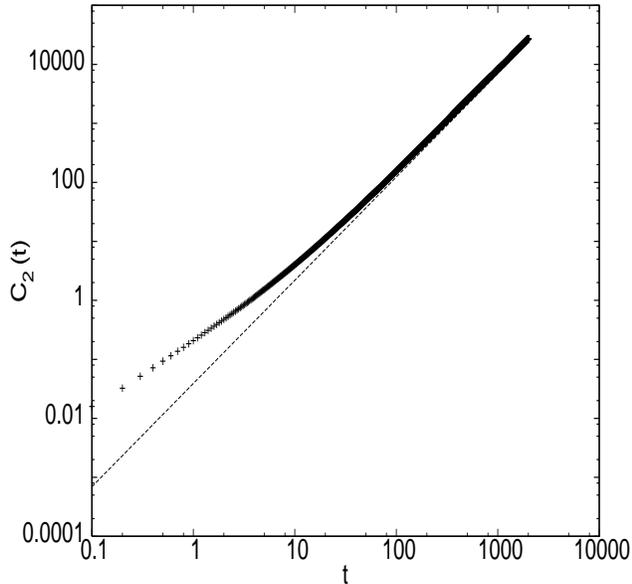}
\caption{Log-log plot of the incremental correlation function $C_2(|t_2-t_1|=t)$ versus
$t$ for the Rouse model. The chain length
is $N = 1024$, while $\Delta t = 0.1$ and $\Delta x = 0.5$; number of
disorder histories $= 30$ and thermal histories $=30$. The numerical
data (shown by $+$ signs) is compared to the theoretical prediction
of a power law with power $7/4$ as shown by the straight line.}
\label{ewC2}
\end{center}
\end{figure}

\begin{figure}
\begin{center}
\epsfxsize=9.0cm \epsfysize=8.0cm
\epsfbox{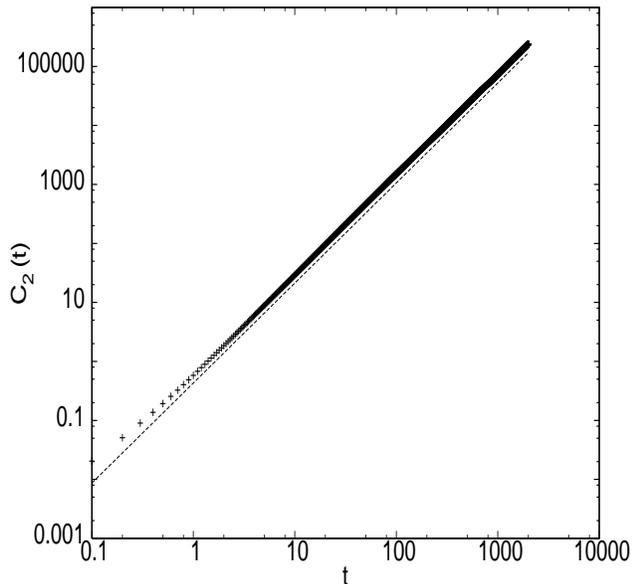}
\caption{Log-log plot of the incremental correlation function $C_2(|t_2-t_1|=t)$ versus
$t$ when $h_1$ evolves via the $(1+1)$-dimensional KPZ equation. The chain length
is $N = 4096$, while $\Delta t = 0.1$ and $\Delta x = 0.5$. The theoretical
prediction is the straight
line that corresponds to a power law in $t$ with a power $5/3\simeq 1.67$, while a straight
line fit to the numerical data (shown by $+$ signs) gives an exponent $\approx 1.695$.
This slight discrepancy is commented upon in the text.}
\label{kpzC2}
\end{center}
\end{figure}

In this Section we numerically verify some of the analytical predictions
made in the previous sections for the coupled interface model, in particular
the fact that in the $t_0\to \infty$ limit  $h_2$ is generically 
a fBm with Hurst exponent $H_2=1-\beta_1/2$, i.e., for large $|t_2-t_1|$,
\begin{equation}
C_2(t_1,t_2, t_0\to \infty)= \lim_{t_0\to \infty} E\left[{\left( h_2({\bf r}, t_0+t_1)-h_2({\bf r}, 
t_0+t_2)\right)}^2\right]\sim |t_2-t_1|^{2-\beta_1} 
\label{c2fbm}
\end{equation}
where $\beta_1$ is the growth exponent of the first surface $h_1$. We have checked
this prediction numerically in $d=1$ for two cases: (i) when the first surface
evolves via the Edwards-Wilkinson equation so that $\beta_1=1/4$ and
Eq. (\ref{c2fbm}) predicts $C_2(t_1,t_2, t_0\to \infty)\sim |t_2-t_1|^{7/4}$ for
large $|t_2-t_1|$. This case
corresponds also to the Rouse chain advected by a transverse velocity field
as mentioned in Section III-B and (ii) when the first surface evolves via
the KPZ equation so that $\beta_1=1/3$. In this case, Eq. (\ref{c2fbm}) predicts
$C_2(t_1,t_2, t_0\to \infty)\sim |t_2-t_1|^{5/3}$. The results for the
simulations in the two cases are shown in Figs. (\ref{ewC2}) and (\ref{kpzC2}) respectively.

Our simulation techniques are straightforward. For the case (i) above we use
the time discretized version of the Rouse chain model, i.e., Eqs. (\ref{evolx1}) and (\ref{evoly1}).
We set $t_m=m {\Delta t}$ and rewrite Eqs. (\ref{evolx1}) and (\ref{evoly1}) as
\begin{eqnarray}
x_n(t_{m+1})&=&x_n(t_m)+ {\Delta t}\left[ x_{n+1}(t_m)+x_{n-1}(t_m)-2\,x_n(t_m)\right]+\sqrt{\Delta t} 
\eta_1(n,t_m) \label{dis1} \\
y_n(t_{m+1})&=& y_n(t_m) + {\Delta t}\left[ y_{n+1}(t_m)+y_{n-1}(t_m)-2\,y_n(t_m)\right] \nonumber \\
&+& {\Delta t}\, v\left(x_n(t_m)\right) + \sqrt{\Delta t}\,\eta_2(n,t_m).
\label{dis2}
\end{eqnarray} 
For the boundary points $n=1$ and $n=N$, we use free boundary conditions, i.e., we hold
$x_0=x_1$, $y_0=y_1$, $x_N=x_{N+1}$ and $y_N=y_{N+1}$ for all times $t_m$.
We choose $\Delta t<0.5$ in our simulations so that the stability is guaranteed\cite{Krug}.
The variables $\eta_1(n, t_m)$ and $\eta_2(n,m)$ are independent Gaussian variables
for all $n$ and $t_m$ and each is distributed with zero mean and unit variance.
We have checked that even if the noise variables have a binary distribution
(i.e., $+1$ and $-1$ each with probability $1/2$), the results at long times 
do not change. Besides, as it turns out from Eq. (\ref{cingen1}) that the thermal
noise $\eta_2$ is actually irrelevant for the long time properties of the $h_2$
process, we have dropped $\eta_2$ in Eq. (\ref{dis2}) in most of our simulations.

We choose the random quenched transverse velocity $v(x)$ in the following way.
First we consider a grid along the $x$ direction with grid spacing $\Delta x$.
In fact this grid represents the layered structure of the velocity field.
At each point of this grid we choose independently $v(x)=u(x)/\sqrt{\Delta x}$ where
$u(x)$ is a Guassian random 
variable with zero mean and unit variance. Once a set of $\{v(x)\}$ is thus chosen, they
remain fixed at all times during different thermal histories. This set $\{v(x)\}$
constitutes a particular realization of disorder. Finally one performs
the disorder average ${\overline {\left(\dots \right)}}$ by averaging over various realizations
of the set $\{v(x)\}$. Now, $x_n$ in Eq. (\ref{dis1}) at any given time
is usually a continuous variable and may not correspond to a grid point. In fact,
in general, $x_n$ will be between two grid points say $x_0$ and $x_0+\Delta x$. 
In such a case, we use, as a convention, $v(x_0)$ while simulating the r.h.s of Eq. (\ref{dis2}).
For a fixed
realization of the disorder $\{v(x)\}$, we average over $30-40$ thermal histories
(generated via $\eta_1$) and then a final average is done over $30-40$ disorder
realizations.   

Using this discretization scheme we have computed the incremental correlation function $C_2(t_1,t_2,t_0)$ as 
defined in Eq. (\ref{c2fbm}) beyond some large waiting time $t_0$ (typically $t_0\sim 30000$ steps)
for different choices of time step $\Delta t$ and the grid size $\Delta x$. We have checked
that the results do not vary much with the time step or the grid size. We find that, as predicted
analytically, $C_2(t_1, t_2, t_0\to \infty)$ depends only on the time difference $|t_2-t_1|$
and for large $t_2-t_1|$, $C_2$ scales as a power law with an exponent $\approx 1.75$ which
is consistent with the theoretical prediction $7/4$. A plot of $C_2$ as a function
of $|t_2-t_1|$ is shown in Fig. (\ref{ewC2}). This confirms numerically the theoretical
prediction that indeed $h_2$
is a fBm with the Hurst exponent $H_2=7/8$ and hence also confirms that the persistence
exponent $\theta_s^2=1/8$.

We also numerically computed the incremental correlation function $C_2(t_1,t_2, t_0\to \infty)$
for case (ii), i.e., when $h_1$ evolves by the $(1+1)$-dimensional KPZ equation.
For the $(1+1)$-dimensional KPZ equation in Eq. (\ref{h1kpz}), one needs to be 
careful about the discretization of space and time as has been discussed 
extensively in the literature\cite{DDK,NB,NewSwi,LS,MC}. In this paper we used
the discretization scheme proposed by Newman and Swift\cite{NewSwi} with periodic boundary condition
\begin{equation}
h_1(x_n,t_{m+1})= {\rm max}\left[ h_1(x_{n+1}, t_m), h_1(x_{n-1}, t_m), h_1(x_n,t_m)\right]+ \sqrt{\Delta t} \, 
\eta_1(x_n, t_m).
\label{nsd}
\end{equation}
We show in Fig. (\ref{kpzC2}) the incremental correlation function $C_2(|t_2-t_1|=t, t_0\to \infty)$
versus $t$. Evidently $C_2(t)$ increases as a power law with an exponent $\approx 1.695$ which 
is close to the theoretical prediction $5/3\simeq 1.67$. The slight discrepancy is due to the system size 
used by
us, for which even the scaling exponent $\beta_1$ for $h_1$
actually is not exactly $1/3$ but close to $\approx 0.305$. We
have checked that the exponent systematically approaches
the expected value with increasing $N$, and here our quoted
value is based on the largest $N$ that we could simulate.

\section{Conclusion}

In this paper we have studied the time-dependent properties in a simple model of coupled interfaces
characretized by heights $h_1$ and $h_2$ respectively growing over a $d$-dimensional
substrate. The evolution of the first interface $h_1$ is not affected by $h_2$. In fact,
$h_1$ can be any generic growing interface characterized by a growth exponent $\beta_1$.
For example, $h_1$ may be evolving either via the Edwards-Wilkinson equation or via
the Kardar-Parisi-Zhang equation. The evolution of the second interface $h_2$ however
is coupled to that of $h_1$ by a transverse quenched random velocity field,
in addition to having the usual Edwards-Wilkinson dynamics. 
In the limiting case $d\to 0$, our model reduces to the 
Matheron-de Marsily model where one studies the motion
of a Brownian particle in a $2$-d plane in presence of
a transverse velocity field. In the limit $d=1$, our model
describes the equations of motion of a Rouse polymer chain
in two dimensions in presence of a transverse velocity field. 

We have obtained analytical results for the persistence properties in this model.
The main result of this paper is to show analytically that after a long waiting time ($t_0\to \infty$),
the process $h_2$, at a fixed point in space but as a function of time, becomes a
fractional Brownian motion with a Hurst exponent $H_2=1-\beta_1/2$. By using a known
first-passage property of fractional Brownian motion we have then shown that 
after a long waiting time $t_0\to \infty$, the persistence probability $P_2(t)$ 
that the process $h_2$ at a fixed point in space does not come back to
its value at $t_0$ over the time interval $[t_0,t_0+t]$ decays as a
power law for large $t$, $P_2(t)\sim t^{-\theta_s^2}$ where the exponent
$\theta_s^2=1-H_2=\beta_1/2$. For $d=1$, these analytical predictions
have been verified numerically in two cases: when $h_1$ evolves
via the Edwards-Wilkinson equation and when $h_1$ evolves via the
Kardar-Parisi-Zhang equation.

The mapping of a relevant stochastic process in some limits to
a fractional Brownian motion is a rather powerful technique for
studying the first-passage properties for non-Gaussian and/or
non-Markovian processes. The mapping does not work
always, but if it works one can use the known first-passage property
of the fractional Brownian motion.
This technique
has been used successfully in a number of contexts 
previously\cite{SM,Krug,Krug1,MB1,M3,Krug2}. We have demonstrated 
that the same technique also works
in another class of coupled interface models discussed in
this paper. It would be interesting to find other such
cases where one can apply the same technique successfully. 
Finally, it would be interesting to study the persistence
properties in more realistic models of coupled interfaces
that are closer to the experimental situation of fluctuating
steps\cite{PE}.

\end{document}